\documentclass[pre,twocolumn,amsmath,amssymb,aps]{revtex4}

\usepackage{graphicx}

\begin{document}

\title{On the role of chemical synapses in coupled neurons with noise}

\author{Pablo Balenzuela}
\email{pablo.balenzuela@upc.edu} \altaffiliation[also at ]{Departamento de F\'isica, FCEyN, Universidad de Buenos Aires, Pabell\'on 1, Ciudad Universitaria (1428), Buenos Aires,
Argentina.}
\author{Jordi Garc\'ia-Ojalvo}
\email{jordi.g.ojalvo@upc.edu}
\affiliation{Departament de F\'isica i Enginyeria Nuclear, Universitat Polit\`ecnica de Catalunya, Colom 11,
E-08222 Terrassa, Spain}

\date{\today}

\begin{abstract}
We examine the behavior in the presence of noise of an array of Morris-Lecar neurons coupled via 
chemical synapses. Special attention is devoted to comparing this behavior with the better known 
case of electrical coupling arising via gap junctions. In particular, our numerical simulations 
show that chemical synapses are more efficient than gap junctions in enhancing coherence at an
optimal noise (what is known as array-enhanced coherence resonance): in the case of (nonlinear) 
chemical coupling, we observe a substantial increase in the stochastic coherence of the system, 
in comparison with (linear) electrical coupling. We interpret this qualitative difference between 
both types of coupling as arising from the fact that chemical synapses only act while the 
presynaptic neuron is spiking, whereas gap junctions connect the voltage of the two neurons at 
all times. This leads in the electrical coupling case to larger correlations during interspike time 
intervals which are detrimental to the array-enhanced coherence effect. Finally, we report on the 
existence of a system-size coherence resonance in this locally coupled system, exhibited by the 
average membrane potential of the array.                                                                                  
\end{abstract}


\maketitle

\noindent {\bf  
}

\section{\label{sec:intro}Introduction}

Neurons are excitable devices that respond in a spiky manner to extrinsec stimuli. These stimuli 
can be provided by external excitation, by noise, or by neighboring neurons in an extended 
system \cite{physrep}. In the absence of deterministic external driving, {\em isolated} neurons 
exhibit a spiking behavior purely induced by noise, with the peculiarity that the temporal 
coherence of the system increases for increasing noise up to a certain noise level, beyond which 
coherence decreases again. Thus, an optimal amount of noise exists for which coherence is maximal. 
This phenomenon, which we call {\em stochastic coherence} (to stress the analogy with the better 
known stochastic resonance), is known in the literature as coherence resonance or internal 
stochastic resonance \cite{piko,lindschi}.

Recent studies have shown that, in extended arrays of neurons, coupling noticeably enhances the 
stochastic coherence effect \cite{dima,changsong}. This {\em array-enhanced stochastic coherence} 
(AESC) has been reported so far, up to our knowledge, only in the case of linear (diffusive) 
electrical coupling, mediated by gap junctions between the neurons \cite{keener}. But another very 
important means of signal transmission between neurons is via chemical synapses, which provide a 
nonlinear pulsed coupling only when the presynaptic neuron is excited. It is thus of interest to 
examine the effect of this kind of nonlinear coupling on the stochastic resonance effect described 
above. Our numerical results, detailed below, show that chemical synapses are more efficient 
at enhancing coherence than gap junctions. We provide a qualitative explanation for this fact, 
paying particular attention to the roles of synchronization and of the correlation between neural dynamics
in the time lapse between spikes. To that end, we examine the effect of a linear pulsed coupling designed 
{\em ad hoc} for this purpose, which exhibits the optimal features of chemical coupling while 
still acting linearly on the membrane potential during spiking.

\section{Model description}

\subsection{Neuron model}

We consider a one-dimensional array of neurons whose dynamical behavior is described by the Morris-Lecar model \cite{morlec},
\begin{eqnarray}
\frac{dV_i}{dt} & = & \frac{1}{C_m}(I^{\rm app}_i - I^{\rm ion}_i-I^{\rm syn}_i)+D_i\xi(t) \label{eqV} \\
\frac{dW_i}{dt} & = & \phi \Lambda(V_i)[W_{\infty}(V_i) - W_i]   \label{eqW}
\end{eqnarray}
where $i=1,\ldots N$ index the neurons, and $V_i$ and $W_i$ represent the membrane potential and the fraction of open potasium channels, respectively. $C_m$ is the membrane capacitance per unit area, $I^{\rm app}_i$ is the external
applied current, $I^{\rm syn}_i$ is the synaptic current, and the ionic current is given by
\begin{eqnarray}
&&I^{\rm ion}_i  =  g_{Ca}M_{\infty}(V_i)(V_i-V^0_{Ca})+ \nonumber \\
 & & \qquad\qquad\qquad g_KW_i(V_i-V^0_K)+g_L(V_i-V^0_L) \label{eqIion}
\end{eqnarray}
where $g_a$ ($a=Ca,K,L$) are the conductances and $V^0_a$ the resting potentials of the calcium, potassium and
leaking channels, respectively. $\phi$ is the decay rate of $W_i$, and we define the following functions of the membrane potential:
\begin{eqnarray}
&& M_{\infty}(V) =  \frac{1}{2}\left[1+\tanh\left(\frac{V-V_{M1}}{V_{M2}}\right)\right]      \\
&&W_{\infty}(V) = \frac{1}{2}\left[1+\tanh\left(\frac{V-V_{W1}}{V_{W2}}\right)\right]      \\
&&\Lambda(V) = \cosh\left(\frac{V-V_{W1}}{2V_{W2}}\right)\,,
\end{eqnarray}
where $V_{M1}$, $V_{M2}$, $V_{W1}$ and $V_{W2}$ are constants to be specified later. The last term in Eq.
(\ref{eqV}) is a white Gaussian noise term of zero mean and amplitude $D_i$, uncorrelated between different neurons.

In the absence of noise, an isolated Morris-Lecar neuron shows a bifurcation to a limit cycle for increasing applied current $I^{\rm app}$ \cite{bifutrue}. This bifurcation can be a saddle-node (type I) or a 
subcritical Hopf (type II) bifurcation, depending on the parameters. We chose this last option for the 
numerical calculations presented in this paper. The specific values of the parameters used are shown in table \ref{tab:ML} \cite{bifu}.
The equations were integrated using the Heun method \cite{nises}, which is a second order Runge-Kutta algorithm for stochastic equations.

\subsection{Coupling scenarios}

Most of the studies done so far in the field of stochastic neural dynamics consider linear electrical coupling through gap junctions between adjacent neurons
\cite{changsong,tseva,shura,moon,bambihu,vibr}. But the most common way used by neurons to transmit information is by means of nonlinear pulsed coupling through chemical synapses. In the following paragraphs we discuss the modeling of the two types of coupling, both of which will be analyzed later.

\subsubsection{\label{subsec:lin} Linear diffusive coupling: gap junctions}

In this kind of coupling, the synaptic current is proportional to the membrane potential difference between a neuron and its neighbors,
\begin{equation}
I^{\rm syn}_i = \sum_{j\in {\rm neigh}(i)} g_{ij}^{\rm syn}(V_i-V_j), \label{linsyn}
\end{equation}
where $V_i$ stands for the membrane potential of neuron $i$, the sum runs over the neighbors that feed that neuron, and $g^{\rm syn}_{ij}$ is the conductance of the synaptic channel.

\subsubsection{\label{subsec:quim} Nonlinear pulsed coupling: chemical synapses}

In order of take into account the chemical nature of the synapses, we use the model 
proposed in \cite{destexhe} to couple the neurons. 
In this model, the synaptic current through neuron $i$ is given by
\begin{equation}
I^{\rm syn}_i = \sum_{j\in {\rm neigh}(i)} g_i^{\rm syn}r_j(V_i-E_s), \label{syn}
\end{equation}
where the sum runs over the neighbors that feed neuron $i$, $g^{\rm syn}_i$ is the conductance of the synaptic channel,
$r_j$ represents the fraction of bound receptors, $V_i$ is the
postsynaptic membrane potential, and $E_s$ is a parameter whose value determines the type of synapse (if larger
than the rest potential, {\em e.g.} $E_s=0$~mV, the synapse is excitatory; if smaller, {\em e.g.} $E_s=-80$~mV,
it is inhibitory).

The fraction of bound receptors, $r_j$, follows the equation
\begin{equation}
\label{receptor}
\frac{dr_j}{dt}= \alpha [T]_j(1-r_j) - \beta r_j\,,
\end{equation}
where $[T]_j=\theta(T_0^j +\tau_{\rm syn}-t)\theta(t-T_0^j)$ is the concentration of neurotransmitter released
into the synaptic cleft, $\alpha$ and $\beta$ are rise and decay time constants, respectively, and $T_0^j$ is the
time at which the presynaptic neuron $j$ fires, which happens whenever the presynaptic membrane potential exceeds a predetermined threshold value, in our case chosen to be $10$~mV. This thresholding mechanism lies at the origin of the nonlinear character of the chemical synaptic coupling, which contrasts with the linear nature of the diffusive electrical coupling of Eq.~(\ref{linsyn}).
The time during which the synaptic connection is active is given by $\tau_{\rm
syn}$. The values of the coupling parameters that we use \cite{destexhe} are specified in Table \ref{tab:ML} . 

\begin{table}[htbp]
\begin{center}
\begin{minipage}[t]{2.5in}
\centering
\begin{tabular}{|c|c|}
 \hline
  \textbf{Parameter}  & \textbf{Value} \\ \hline
 $C_m$  &  $5\,\mu \mathrm{F/cm}^2$    \\ \hline
 $g_K$  &  $8\,\mu \mathrm{S/cm}^2$   \\  \hline
 $g_L$  &  $2\,\mu \mathrm{S/cm}^2$     \\ \hline
 $g_{Ca}$  & $4.4\,\mu \mathrm{S/cm}^2$    \\   \hline
 $V_K$  &  $-80\,\mathrm{mV}$     \\ \hline
 $V_L$  &  $-60\,\mathrm{mV}$     \\ \hline
 $V_{Ca}$ & $120\,\mathrm{mV}$     \\ \hline
 $V_{M1}$  &  $-1.2\,\mathrm{mV}$     \\ \hline
 $V_{M2}$  &  $18\,\mathrm{mV}$     \\ \hline
 $V_{W1}$  &  $2\,\mathrm{mV}$     \\ \hline
 $V_{W2}$  &  $30\,\mathrm{mV}$     \\ \hline
 $\phi$  & $1/25\, \mathrm{s}^{-1}$     \\ \hline
  $\alpha$  & $0.5\, \mathrm{ms}^{-1} \mathrm{mM}^{-1}$ 	\\ \hline
 $\beta$   & $0.1\, \mathrm{ms}^{-1}$ 	\\ \hline
 $g^{\rm syn}$ & (specified in each case)	\\ \hline
 $\tau_{\rm syn}$ & (specified in each case)  \\ \hline
 $E_s$ & $0\,\mathrm{mV}$	\\ \hline
\end{tabular}
\caption{Parameters used in this work.\label{tab:ML}}
\end{minipage}
\end{center}
\end{table}
		
\section{The behavior of $N$ coupled neurons: Chemical synapses vs gap junctions}

To obtain a first glimpse of the different effects of chemical and electrical synapses in neural 
dynamics, we begin by studying the behavior of {\em two} coupled neurons under the influence of 
increasing noise intensity $D$, for the two types of coupling.
We fix the external current for the two neurons at $I^{\rm app}=46$~mA, in such a way that they do 
not fire in absence of noise. The conductance in the synaptic channels (a measure of the coupling 
strength) was chosen so that the neurons fire synchronously at least for low noise intensities 
($g^{\rm syn}=4$~nS for chemical synapses and $g^{\rm syn}=1$~nS for electrical coupling). 
We use independent realizations of the noise in each neuron but with the same amplitude $D$.

\begin{figure}
\includegraphics[height=5.8cm,keepaspectratio,clip]{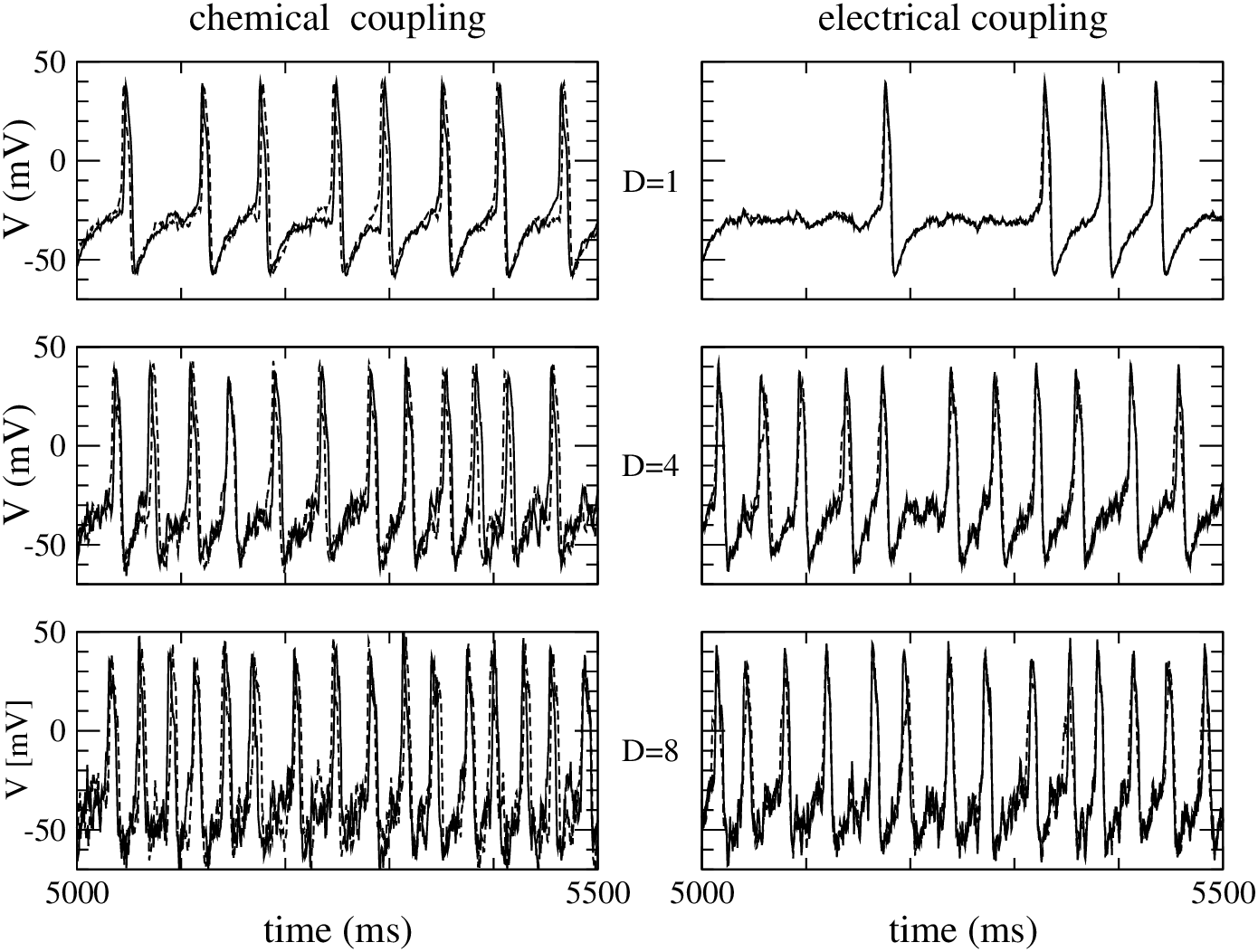}
\caption{\label{fig:1Nvt} Membrane potential time traces of two neurons coupled via chemical synapses (left panels) and gap junctions (right panels) for three differents noise amplitudes:
in the chemical-coupling case $D=0.5$~mV/ms (top), $D=3.5$~mV/ms (middle) and $D=12$~mV/ms (bottom);
in the electrical-coupling case $D=1$~mV/ms (top), $D=4$~mV/ms (middle) and $D=12$~mV/ms (bottom). 
The time series of both neurons are shown as solid and dashed lines (almost indistinguishable in 
the top right panel). $I^{\rm app}=46$~mA, and $\tau_{\rm syn}=1.5$~ms in all cases. 
For chemical coupling, $g^{\rm syn}=4$~nS (left), whereas for the electrical coupling
$g^{\rm syn}=1$~nS (right).}
\end{figure}

In the deterministic case ($D=0$) the system is in a quiescent state. At small but nonzero noise 
levels the neurons spike sparsely (albeit synchronously), producing an irregular sequence of 
well-separated spikes, as can be seen in the top row of Fig.~\ref{fig:1Nvt}. For an intermediate 
noise amplitude the spiking rate increases, as does the regularity of the spiking events 
(middle row in Fig.~\ref{fig:1Nvt}). Finally, for strong noise the spiking of the neurons becomes 
irregular again (bottom row in Fig.~\ref{fig:1Nvt}). The situation is qualitatively the same for 
both types of coupling, even though when comparing the most optimal situation in each case, the 
coherence in the case of chemical coupling is larger than for electrical coupling (as we quantify 
below). We note that even though the two neurons fire synchronously in both cases, in the 
chemical-coupling case there exists a slight delay between spikes, and the inter-spike dynamics 
are different from each other ({\em cf} the solid and 
dashed lines in all plots of Fig.~\ref{fig:1Nvt}). In the electrical-coupling case, on the other 
hand, the time traces are basically identical at all times. This is our first indication that gap 
junctions are much more efficient than chemical synapses in leading to synchronization.

As usual in neurophysiology, in order to quantify the behavior shown above we evaluate the time 
interval between consecutive spikes, $T_p$, as the main observable in our numerical simulations. 
In particular, we analyze the first two statistical moments of the distribution of $T_p$, namely 
its mean value $\langle T_p\rangle $ and its normalized standard deviation (also known as 
coefficient of variation) $R_p = \sigma_p/\langle T_p\rangle $.

\begin{figure}
\includegraphics[height=5.1cm,keepaspectratio,clip]{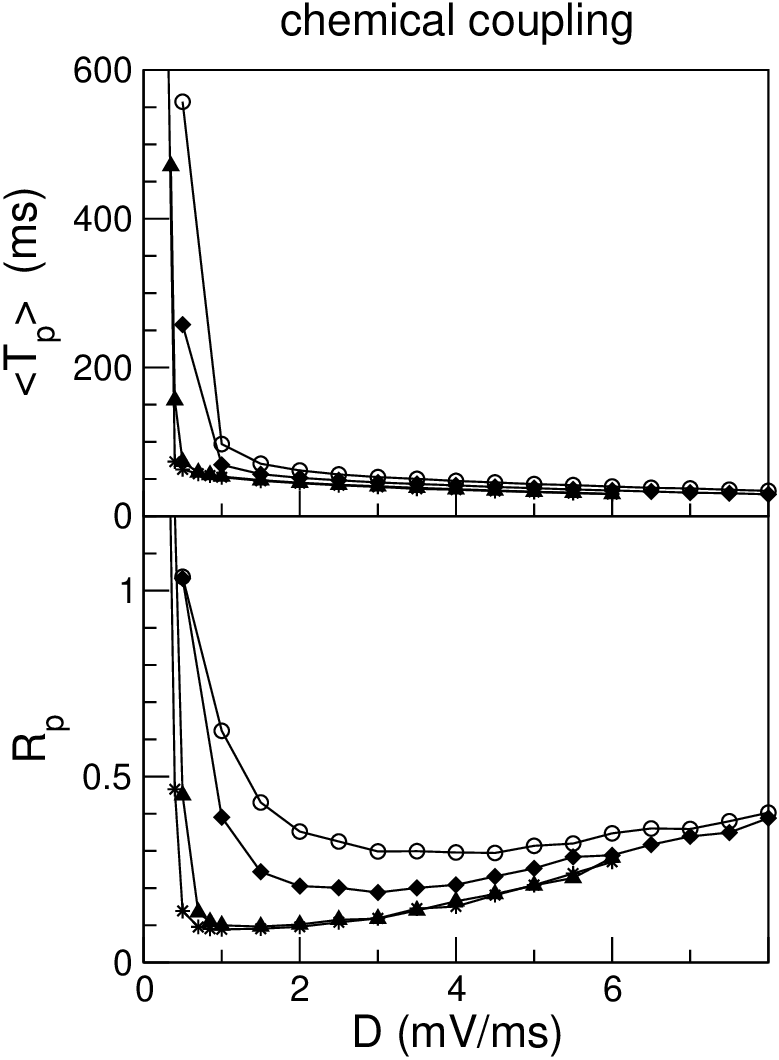}
\hskip1mm
\includegraphics[height=5.1cm,keepaspectratio,clip]{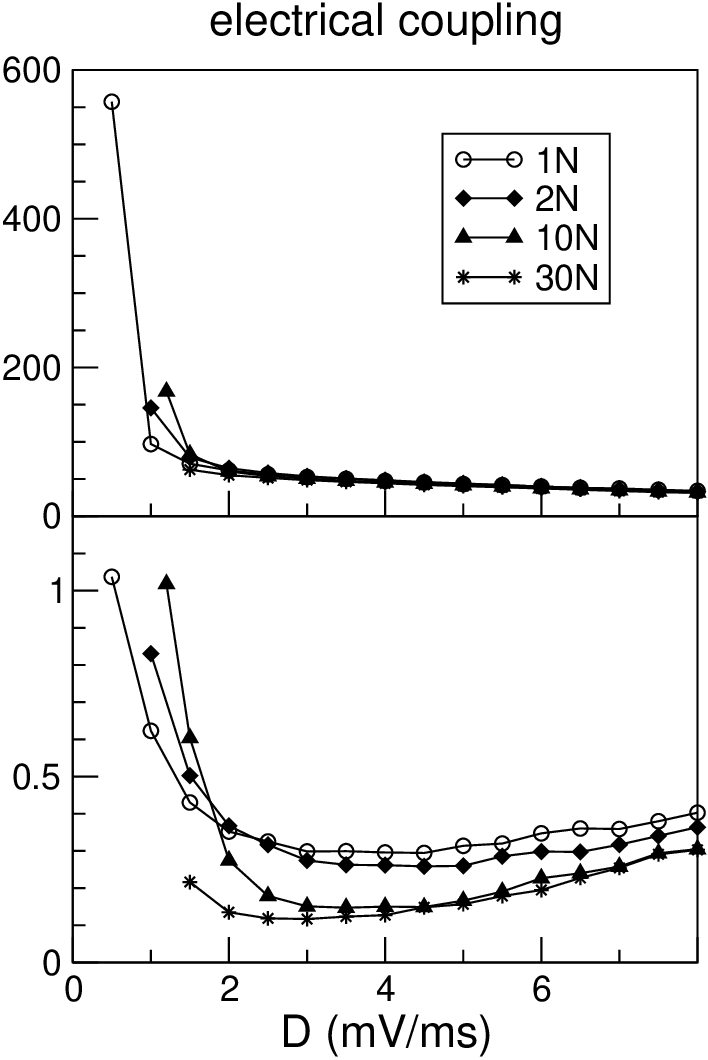}
\caption{\label{fig2} Mean time interval between spikes $\langle T_p\rangle $ (upper panels) and 
coefficient of variation $R_p=\sigma_p/\langle T_p\rangle $ (lower panels) for the 
membrane potential of one neuron coupled via chemical synapses (left panels) and gap junctions 
(right panels). $I_{app}=46$~mA and $\tau_{\rm syn}=1.5$~ms in all cases, whereas 
$g^{\rm syn}=4$~nS for the chemical coupling (left), and $g^{\rm syn}=1$~nS for the electrical 
coupling (right).}
\end{figure}

Figure~\ref{fig2} plots these two quantities vs. noise strength for both chemical and electrical 
coupling in linear arrays of $N=1, 2, 10$ and $30$ neurons, coupled bidirectionally and with 
periodic boundary conditions. While for both types of coupling the average interspike interval 
$\langle T_p\rangle $ decays quickly with noise and levels off independently of the system size, 
the coefficient of variation shows clear differences between chemical and electrical coupling. 
In the former case, $R_p$ shows a clear trend as $N$ varies, with both its minimum value and the 
corresponding optimal noise level decreasing steadily (and strongly) with increasing $N$. For 
electrical coupling, on the other case, the decrease is less pronounced. For a given system size, 
the coherence of the chemically coupled array is much better than that of the electrically coupled 
array.

\begin{figure}
\includegraphics[width=0.35\textwidth,keepaspectratio,clip]{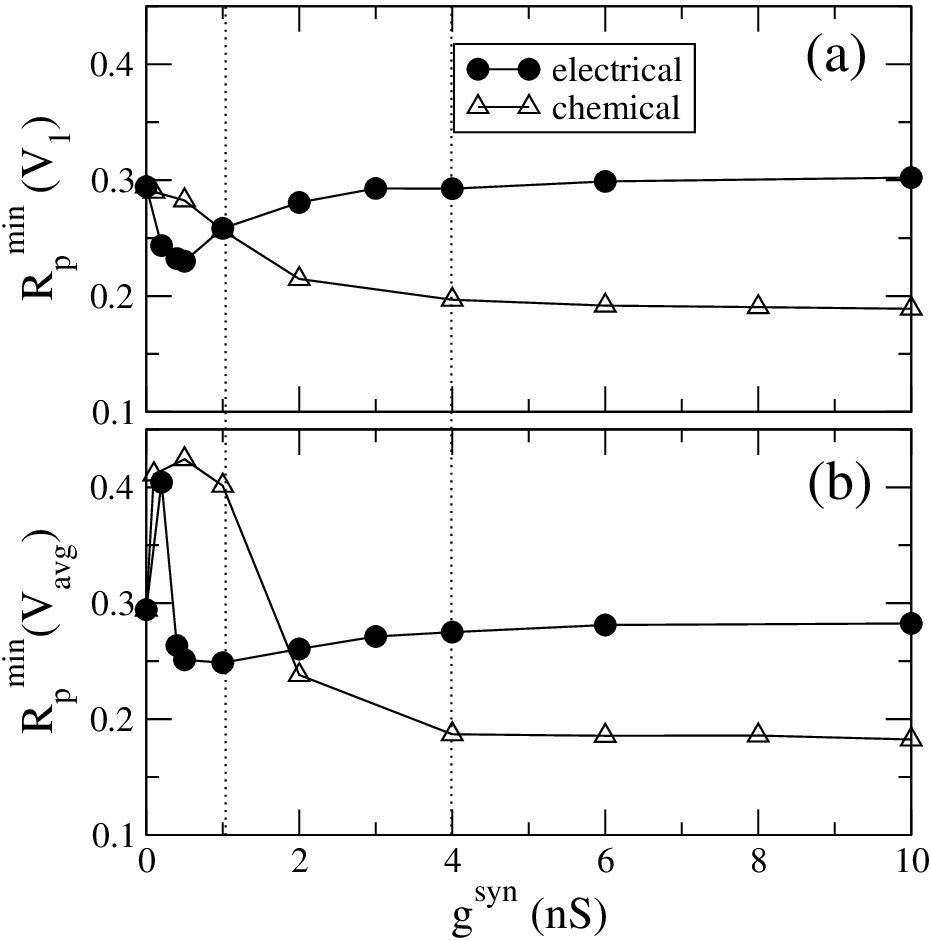}
\caption{\label{newfig3} Minimum coefficient of variation $R_p=\sigma_p/\langle T_p\rangle$ of the local (top) and average (bottom) membrane potential for two coupled neurons versus coupling strength, for both chemical and electrical coupling. $I_{app}=46$~mA in all cases.}
\end{figure}

It could be argued that the difference in behavior between the two types of coupling is due to the 
different coupling coefficients $g^{\rm syn}$ used. But an analysis of the influence of this 
parameter in each model confirms that chemical coupling is overall more efficient than the 
electrical one in enhancing coherence. This is evidenced by Fig.~\ref{newfig3}(a), which plots the 
minimum coefficient of variation $R_p$ ({\em i.e.} its value for optimal noise) versus the coupling coefficient $g^{\rm syn}$ for the two
types of coupling in the 2-neuron case. It is clear that optimal coherence under electrical 
coupling is only attainable for a narrow range of coupling strengths. For small and large coupling 
levels coherence is low, and only for a small intermediate range of coupling strengths coherence is
enhanced. On the other hand, for chemical coupling the range of coupling strengths for which 
coherence is enhanced is much larger, apparently not being bounded from above.

\section{Synchronization and array enhancement}

When choosing the right values of the coupling coefficient that make chemical and electrical coupling comparable, one also needs to take into account the synchronization between the different elements in the array. This can be accomplished via the average 
membrane potential $V_{\rm avg}$ over all the array. In the case of perfect synchronization, this 
quantity will exhibit spikes identical to, and simultaneous with, those of the individual neurons. 
In the case of partial synchronization, spikes in the average potential (defined beyond a given 
threshold, in our case $10$~mV) only occur when a sufficient number of neurons fire within a time 
window smaller than the spike width. Given these considerations, we determine the interspike 
interval series of the average potential and compute its statistical properties, namely its mean 
$\langle T_p\rangle $ and its normalized standard deviation $R_p = \sigma_p/\langle T_p\rangle$. 

Figure \ref{newfig3}(b) plots the minimum value of $R_p$ (for optimal noise) corresponding to the average potential, for increasing coupling strengths. On the basis of these results, we can establish that the best choices for coupling strengths are $g^{\rm syn}=1$~nS for electrical coupling and 
$g^{\rm syn} \ge 4$~nS for chemical. These values are valid not only for two neurons, but for all analyzed system sizes.

We have seen so far that chemical coupling is more effective at enhancing 
stochastic coherence than electrical coupling. We now discuss the 
relationship between array-enhanced coherence and synchronization. 
Array-enhanced coherence is absent for low levels of synchronization, because in that case every 
neuron behaves as basically uncoupled from all others, so that no constructive effects of coupling 
(such as array-enhanced coherence) can arise. In the opposite limit, large synchronization levels 
are not useful in promoting coherence either, since in that case the whole system behaves as a 
single neuron. However, in the intermediate regime of imperfect synchronization, neighboring 
neurons can ``remind'' each other to spike at the right times (i.e., right after the refractory 
period) for an optimal noise level, so that coherence is globally enhanced. This cannot happen 
if the neurons are perfectly synchronized.

We now test the efficiency of the two coupling schemes in leading to synchronization. We can 
already expect, from the functional form of the corresponding coupling term, 
Eqs.~(\ref{syn})-(\ref{receptor}), that chemical synapses will never lead to perfect isochronous 
synchronization in neuronal arrays. To begin with, the dynamics of the fraction $r_j$ of bound 
receptors introduces a delay in the interneuronal communication that is absent in the electrical 
coupling case, Eq.~(\ref{linsyn}). The diffusive form of the latter, furthermore, is compatible 
with the identical synchronization solution, $V_i(t)=V_j(t)\;\forall i,j$, 
whereas this is not so for chemical coupling.

\begin{figure}[htb]
\includegraphics[height=5.1cm,keepaspectratio,clip]{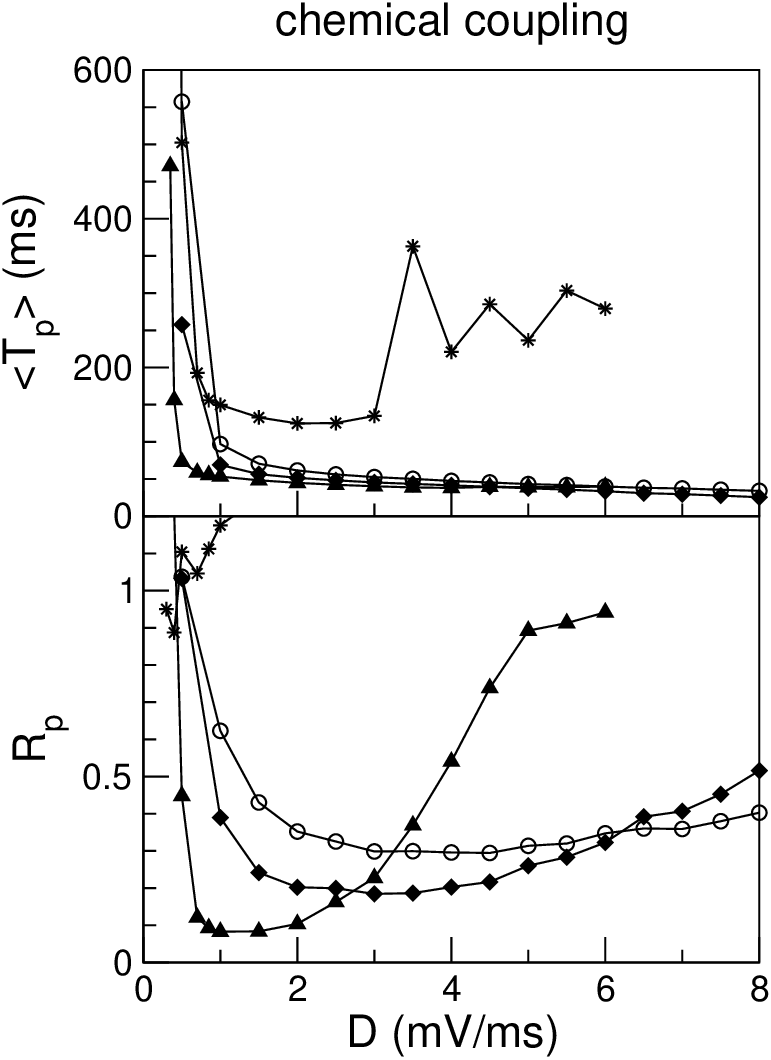}
\hskip1mm
\includegraphics[height=5.1cm,keepaspectratio,clip]{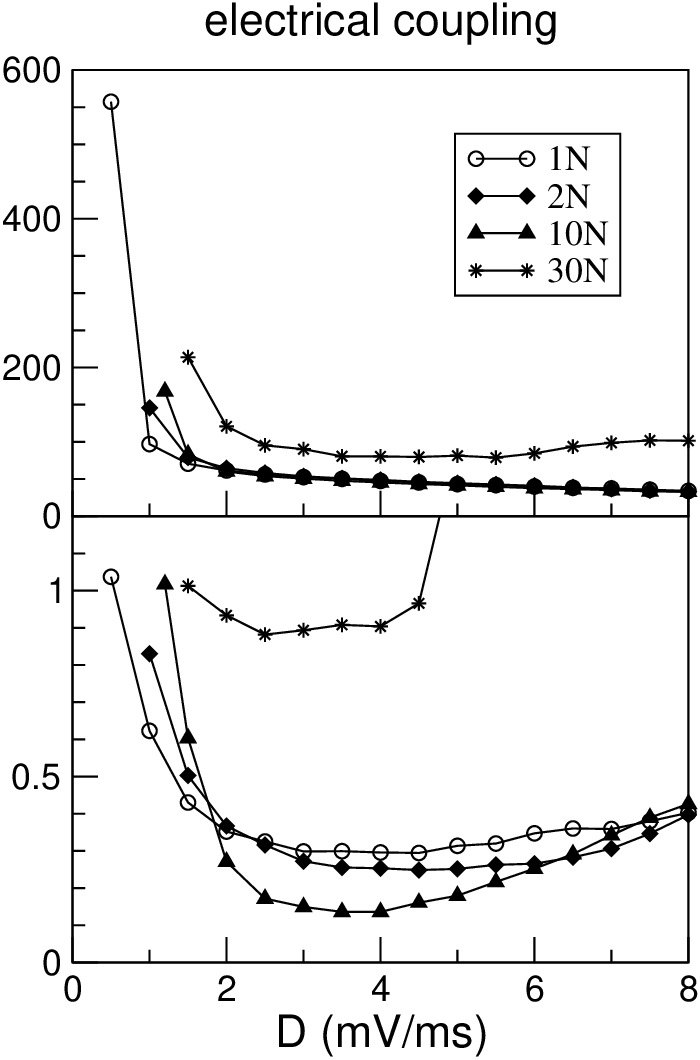}
\caption{\label{fig3} Mean time between spikes $\langle T_p\rangle $ (upper panels) and
coefficient of variation $R_p=\sigma_p/\langle T_p\rangle $ (lower panels) for the
average membrane potential and increasing system sizes. 
Parameters are those of Fig.~\protect\ref{fig2}.}
\end{figure}

To check the previous hypothesis, Fig.~\ref{fig3} compares the mean interspike interval and coefficient of variation of the average potential for the two types of coupling. This figure should 
be contrasted with the corresponding results for the local membrane potential, Fig.~\ref{fig2}. 
Again, the situation is qualitatively similar in both cases, although there are quantitative 
differences. The mean time interval between spikes of the average potential increases noticeably 
for large system sizes ({\em cf} stars in the top row of Fig.~\ref{fig3}), but much more for 
chemical coupling. Also, the coefficient of variation of the interspike interval series becomes 
more degraded  ({\em i.e.} the system response is worst at the same system sizes) in the chemical synapse case. 
These facts indicate that synchronization is worse for chemical than for electrical coupling.

\begin{figure}[htb]
\includegraphics[height=5.8cm,keepaspectratio,clip]{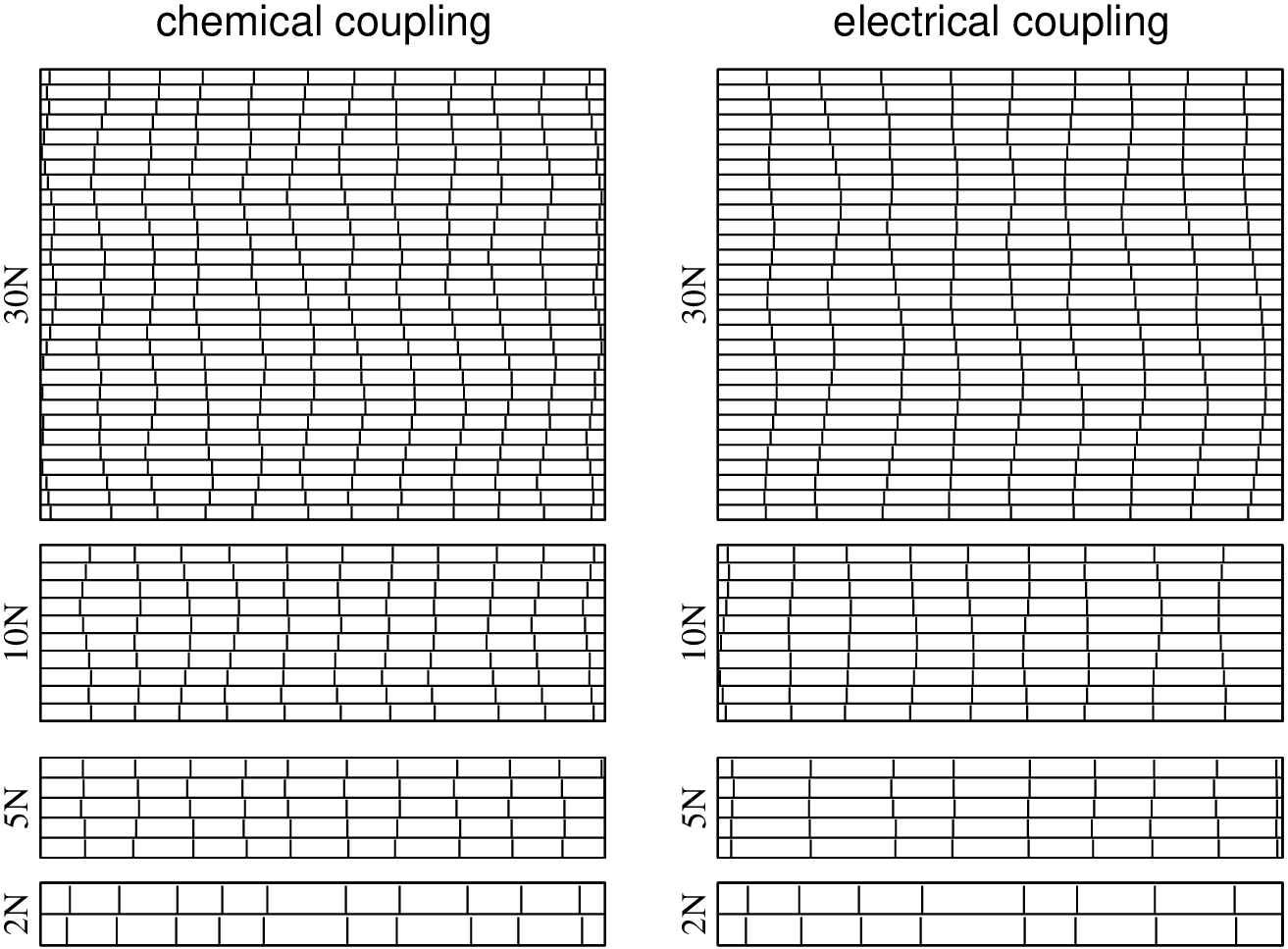}
\caption{\label{fig4} Spike diagram for $N$ neurons coupled via chemical synapses 
($g^{\rm syn}=4$~nS left panels) and gap junctions ($g^{\rm syn}=1$~nS rigth panels) for 
$2$, $5$, $10$ and $30$ neurons. $I_{app}=46$~mA and $D=3$~mV/ms in all cases. 
}
\end{figure}

In order to confirm the previous conclusions, we follow how spikes propagate through the array by 
means of the spike diagram shown in Fig.~\ref{fig4}. In this diagram, we plot  vertical lines 
every time that one neuron spikes. In this kind of diagram synchronized spiking is clearly 
identified by vertical alignment of the spike markers.

Figure~\ref{fig4} plots the spike diagram for the two types of coupling discussed so far and for 
$N=2,5,10$ and $30$ neurons. By comparing the two columns we can observe that the system with 
chemical synapses has a charateristic delay time in the propagation of spikes between neurons 
larger than the corresponding to the electrical coupling. Due to this delay, synchronization is 
rapidly lost in the case of chemical coupling. For electrical coupling the 
spikes also show slight propagation delays, but much smaller than in the chemical case. With the values of 
coupling constants chosen here, both systems lose synchronicity at similar sizes. If we 
increased the value of the $g^{syn}$ in the electrical coupling case, the system would remain 
synchronized for higher system sizes. This would not happen in the chemical case, because there the delay is 
intrinsic to the synapses and does not depend on the coupling strength. This is a fundamental, qualitative difference between
chemical and electrical coupling, which is worth being highlighted.

\section{Nonlinear versus pulsed coupling}

There are two main differences between coupling via chemical synapses and gap junctions. First, 
electrical coupling occurs continuously whereas chemical synapses are only active when the 
presynaptic neuron spikes. Second, chemical coupling is intrinsically nonlinear, because beyond a 
certain threshold for the presynaptic neuron, the coupling signal has always the same shape. 
Which one of these two features leads to enhancement of stochastic coherence reported above for the 
chemical coupling? We postulate that the relevant feature that makes the chemical synapses more 
efficient to help neurons to fire is that chemical coupling is only effective when the presynaptic 
neuron fires. Otherwise, two neighboring neurons are uncoupled.

To test this hypothesis we couple the neurons with an artificial synaptic current proportional to 
the difference between the voltages of the coupled membranes (as in the usual linear coupling 
described in Sect.~\ref{subsec:lin}), but just during a time $\tau_{\rm syn}$ after the presynaptic
neuron has fired. This model, that we can call {\it linear pulsed coupling}, is a hybrid of the 
two models used above: the coupling is linear with the membrane potential, but the neurons are 
uncoupled as far as the presynaptic neuron does not fire.

The synaptic current affecting neuron $i$ due to the interaction with its neighboors 
can be written as
\begin{equation}
I^{\rm syn}_i = \sum_{j\in {\rm neigh}(i)} \theta(T_0^j +\tau_{\rm syn}-t)\theta(t-T_0^j) g_i^{\rm syn}r_j(V_i-V_j), \label{pulsyn}
\end{equation}
where $\theta(T_0^j +\tau_{\rm syn}-t)\theta(t-T_0^j)$ is a pulse of width $\tau_{\rm syn}$ that 
turns on when the membrane potential of neuron $j$ is larger that a certain threshold ($10mV$ in 
the present case). With this model, a neuron is uncoupled from its neighbors if they are silent, 
and coupled during a time $\tau_{\rm syn}$ after one of them fires.

\begin{figure}
\includegraphics[height=5.8cm,keepaspectratio,clip]{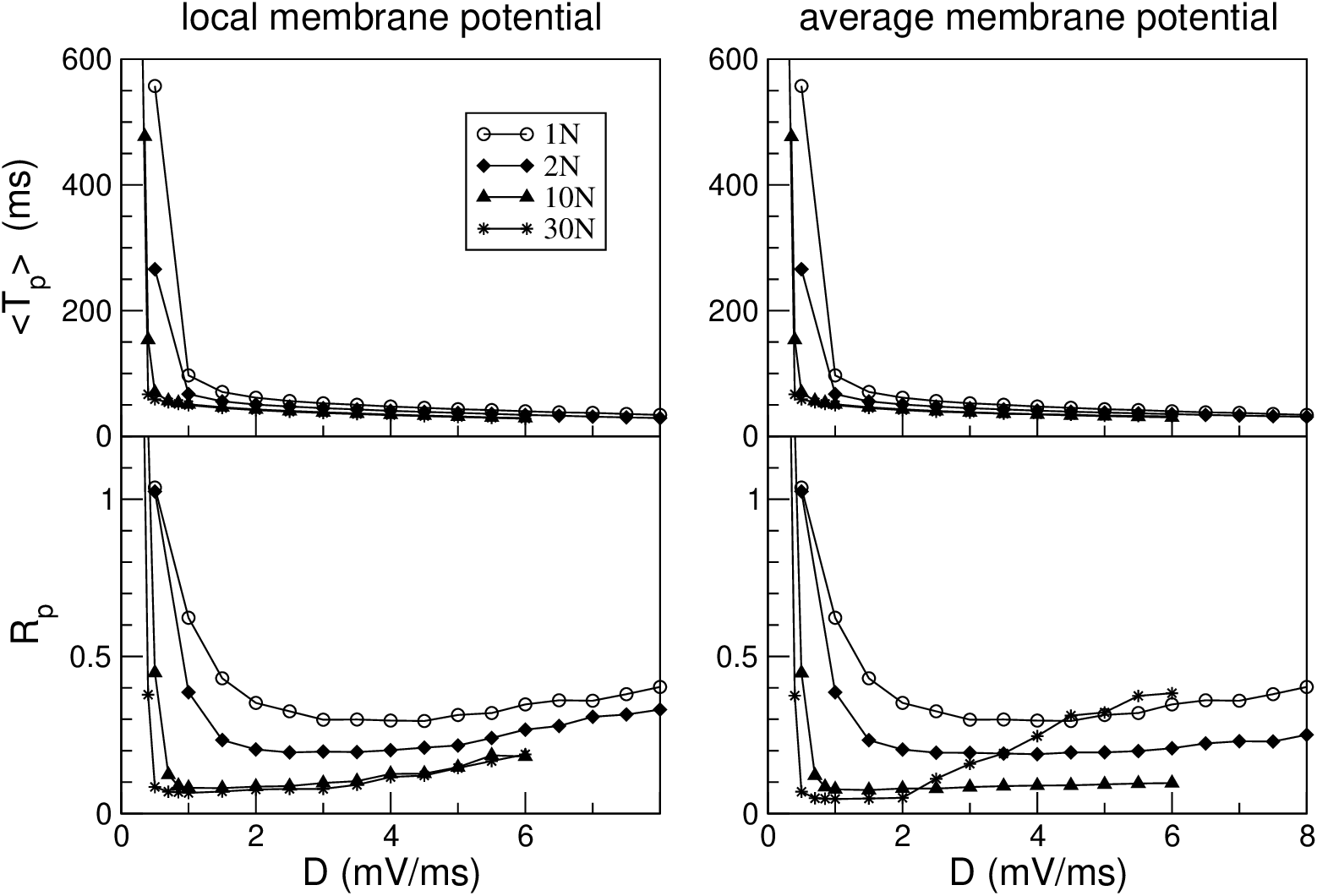}
\caption{\label{fig5} Mean time between spikes $\langle T_p\rangle $ (upper plot) and 
coefficient of variation $R_p=\sigma_p/\langle T_p\rangle $ (lower plot) for the 
membrane potential of one neuron (lef panel) and for the average membrane potential (right panel). 
The neurons are coupled linearly according with expression \protect\ref{pulsyn}. Parameters are 
those of Fig.~\ref{fig2} ($g_{syn}=4$~nS).}
\end{figure}

We also calculate in this case the average time between consecutive spikes, $\langle T_p\rangle $ 
and its normalized standard deviation $R_p = \sigma_p/\langle T_p\rangle$, both for the spikes 
produced by the membrane potential of one neuron and for the average membrane potential. The 
results are plotted in Fig.~\ref{fig5}.

Analyzing the results for the local membrane potential (left panel of Fig.~\ref{fig5}), we can see 
that the behavior in this case is similar to (even better than) the case of full chemical 
coupling: the response of a neuron to a purely noisy excitation is clearly enhanced already when 
only two neurons are coupled, and rapidly improves with system size, saturating after $N=10$ 
neurons. Also, the coefficient of variation $R_p$ is as low as for chemical synapses. This happens 
even though coupling is linear when active. 

If we analyze the behavior of the average potential (right panel of Fig.~\ref{fig5}), we can see that
synchronization holds at even larger system sizes than for the previous natural coupling schemes.
In this hybrid model, the $(V_i-V_j)$ term enhances synchronization for large coupling strengths,
but without losing enhancement as in the linear coupling model, due to the inactivation of the
coupling in the time elapsed between spikes.

In summary, array-enhanced stochastic coherence is present for linear pulsed coupling. The fact that, in this hybrid type of coupling,
this term only turns on when presynaptic neurons fire,
supports the hypothesis that the most important feature underneath the efficiency of 
the coupling is that the neurons remains uncoupled in absence of firing. 

\section{System-size coherence resonance}

The behavior exhibited in Fig.~\ref{fig3} by the coefficient of variation $R_p$ of the average 
membrane potential in the chemical-coupling case displays a relevant feature: for small enough 
noise levels (such as $D=1$~mV/ms), as the system size increases the regularity of the system 
increases ({\em i.e.} the coefficient of variation decreases), only to degrade again for even 
larger system sizes. This is a system-size resonance effect. System-size coherence resonance has 
been recently reported in globally coupled excitable elements \cite{toral} and in arrays of neurons interacting via linear diffusive coupling \cite{wang}. Here we report it for chemically coupled neurons.

\begin{figure}
\includegraphics[height=5.0cm,keepaspectratio,clip]{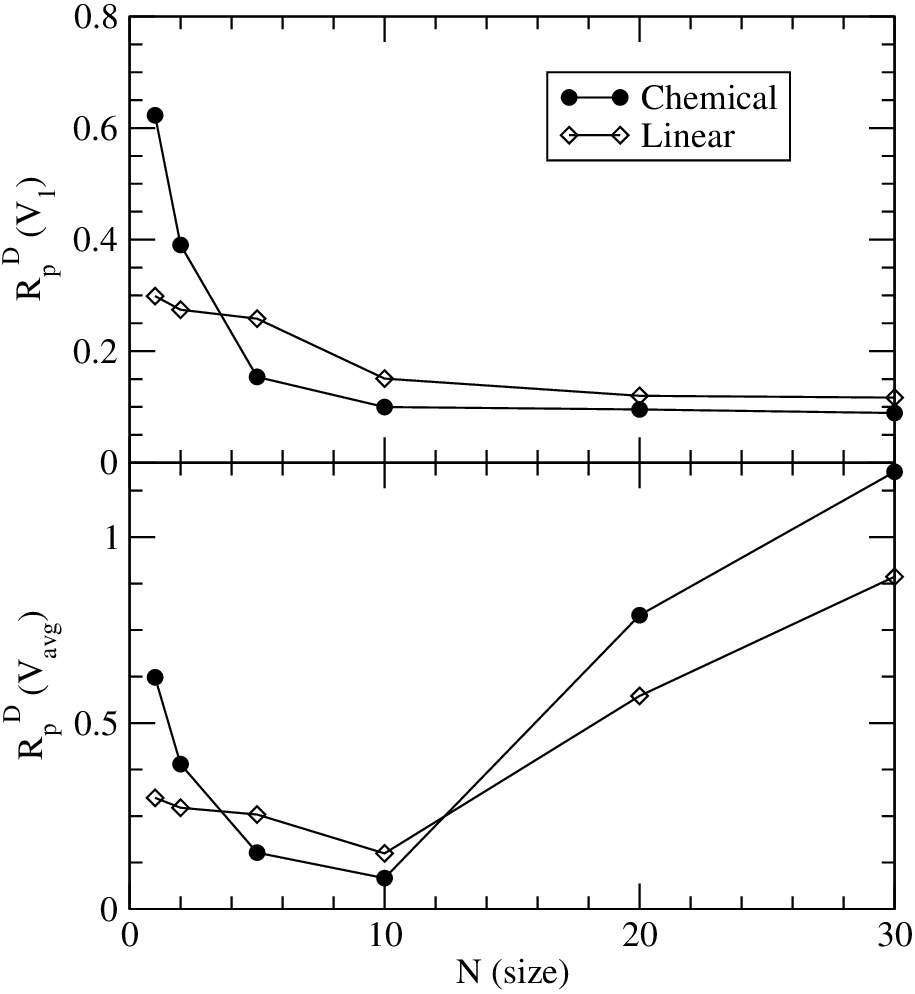}
\caption{\label{fig6} Coefficient of variation $R_p$ for the membrane potential of one neuron as a function of system size for chemical ($D=1$~mV/ms) and electrical ($D=3$~mV/ms) coupling. 
$g^{\rm syn}=4$~nS for chemical coupling and $g^{\rm syn}=1$~nS for the electrical coupling.
}
\end{figure}

Figure~\ref{fig6} compares the effect of system size on the spike-train coherence for chemically 
and electrically coupled neurons. Whereas in both cases, the lowest coherence is found for an 
optimal system size, the effect is more pronounced, and occurs at the same system size 
for the values of $g^{syn}$ chosen.

\section{Discussion}

Recent studies have shown that gap junctions are more efficient than chemical synapses in leading to synchronization
\cite{kopell}. 
The evidence presented in this paper shows that chemical synapses are substantially more efficient 
than gap junctions in enhancing stochastic coherence in coupled neuron systems. 
This difference in efficiency stems from the fact that electrical coupling 
enhances correlations among neurons during interspike time intervals, which prevents 
neighboring elements to ``remind'' each other to fire at the right time.
Chemical coupling, on the other hand, has an intrinsic transmission delay and an intrinsic 
electrical isolation that degrades this correlation, making that cooperative effect possible. 
This is also reflected in the different behavior of both 
types of coupling versus coupling strength (Fig.~\ref{newfig3}): for electrical coupling coherence 
is low only for a narrow window of coupling strengths; for chemical coupling, on the other hand, 
low coherence arises even if coupling is large. Clearly this difference arises from the fact that 
for diffusive coupling, large coupling levels lead to high synchronization; for chemical coupling, 
on the other hand, synchronization does not improve even if coupling strength increases, due to 
the intrinsic delay.

Interneuron communication via chemical synapses is ubiquitous. Its role must therefore be properly 
assessed when studying stochastic effects in neuronal dynamics. The results presented here show 
that this type of coupling is in fact beneficial for the system coherence.

\begin{acknowledgments}
We acknowledge financial support from MCyT-FEDER (Spain, project BFM2003-07850), and from the Generalitat de Catalunya. P.B. acknowledges financial support from the Fundaci\'on Antorchas (Argentina), and from a C-RED grant of the Generalitat de Catalunya.
\end{acknowledgments}

\end{document}